\documentclass[12pt]{article}
\makeatletter
\let\frontmatter@title@above=\relax
\makeatother
\usepackage{times}
\usepackage{geometry}
\usepackage[utf8]{inputenc}
\usepackage{enumitem,amssymb}
\usepackage{ragged2e}
\usepackage{setspace}
\usepackage{graphicx}
\usepackage{floatrow}
\usepackage{orcidlink}
\usepackage{color}
\usepackage{tcolorbox}
\usepackage{xcolor}
\hypersetup{
    colorlinks,
    linkcolor={red!50!black},
    citecolor={blue!50!black},
    urlcolor={blue!80!black}
}
\usepackage{style_aastex}
\newlist{thematic}{itemize}{8}
\setlist[thematic]{label=$\square$}
\usepackage{pifont}

\definecolor{myblue}{HTML}{403BEF}
\definecolor{boxcolor}{HTML}{EFEF3B}
\definecolor{tableblue}{HTML}{1A73C9}
\definecolor{mygrey}{HTML}{363636}
\newtcolorbox{bangbox}[1][]
{
  colframe=white,
  width=\textwidth, 
  colback=boxcolor,
  coltext=black} 

\begin{document}
\pagenumbering{gobble}
\RaggedRight
\noindent {\fontsize{16}{20} \selectfont White Paper for the 2024 Solar \& Space Physics Decadal Survey}
\begin{center}
{\fontsize{22}{32}\selectfont Quantifying Energy Release in Solar Flares \\
\vspace{0.1cm}
and Solar Eruptive Events}
\vspace{0.5cm}

\textit{\fontsize{16}{20}\selectfont New Frontiers with a Next-Generation Solar Radio Facility}
\end{center}


\normalsize

\justifying


\bigskip

\noindent \textbf{Principal Author:} \\
Bin Chen~\orcidlink{0000-0002-0660-3350}, \textit{New Jersey Institute of Technology} \\
Email: \href{mailto:binchen@njit.edu}{binchen@njit.edu}; Phone: (973) 596-3565; Web: \href{https://binchensun.org}{https://binchensun.org}

\smallskip

\noindent \textbf{Co-authors}\\
{\fontsize{12}{14}\selectfont 
Dale~E.~Gary$^{1}$%
\orcidlink{0000-0003-2520-8396},
Sijie~Yu$^{1}$%
\orcidlink{0000-0003-2872-2614}, 
Surajit~Mondal$^{1}$%
\orcidlink{0000-0002-2325-5298},
Gregory~D.~Fleishman$^{1}$%
\orcidlink{0000-0001-5557-2100},
Xiaocan~Li$^{2}$%
\orcidlink{0000-0001-5278-8029},
Chengcai Shen$^{3}$%
\orcidlink{0000-0002-9258-4490}, 
Fan Guo$^{4}$%
\orcidlink{0000-0003-4315-3755}, 
Stephen~M.~White$^{5}$%
\orcidlink{0000-0002-8574-8629},
Timothy~S.~Bastian$^{6}$%
\orcidlink{0000-0002-0713-0604},
Pascal Saint-Hilaire$^{7}$%
\orcidlink{0000-0002-8283-4556},
James F. Drake$^{8}$%
\orcidlink{0000-0002-9150-1841}, 
Joel Dahlin$^{9}$%
\orcidlink{0000-0002-9493-4730},
Lindsay Glesener$^{10}$%
\orcidlink{0000-0001-7092-2703}, 
Hantao Ji$^{11}$%
\orcidlink{0000-0001-9600-9963},
Astrid Veronig$^{12}$%
\orcidlink{0000-0003-2073-002X}, 
Mitsuo Oka$^{7}$%
\orcidlink{0000-0003-2191-1025},
Katharine K. Reeves$^{3}$%
\orcidlink{0000-0002-6903-6832},
Judith Karpen$^{9}$%
\orcidlink{0000-0002-6975-5642}
}

{\fontsize{12}{14}\selectfont \noindent 
[1] New Jersey Institute of Technology; 
[2] Dartmouth College;
[3] Harvard-Smithsonian Center for Astrophysics,
[4] Los Alamos National Laboratory, 
[5] Air Force Research Laboratory,
[6] National Radio Astronomy Observatory,
[7] University of California, Berkeley,
[8] University of Maryland
[9] NASA Goddard Space Flight Center,
[10] University of Minnesota,
[11] Princeton University,
[12] University of Graz
}

\noindent \textbf{Co-Signers \& Affiliations:} see spreadsheet

\vspace{0.2cm}

\noindent \textbf{Synopsis} \\
Solar flares and the often associated solar eruptive events serve as an outstanding laboratory to study the magnetic reconnection and the associated energy release and conversion processes under plasma conditions difficult to reproduce in the laboratory, and with considerable spatiotemporal details not possible elsewhere in the universe. In the past decade, thanks to advances in multi-wavelength imaging spectroscopy, as well as developments in theories and numerical modeling, significant progress has been made in improving our understanding of solar flare/eruption energy release. In particular, \textit{broadband imaging spectroscopy at microwave wavelengths} offered by the Expanded Owens Valley Solar Array (EOVSA) has enabled the revolutionary capability of measuring the time-evolving coronal magnetic fields at or near the flare reconnection region. However, owing to EOVSA's limited dynamic range, imaging fidelity, and angular resolution, such measurements can only be done in a region around the brightest source(s) where the signal-to-noise is sufficiently large. In this white paper, after a brief introduction to the outstanding questions and challenges pertinent to magnetic energy release in solar flares and eruptions, we will demonstrate how a next-generation radio facility with many ($\sim$100--200) antenna elements can bring the next revolution by enabling high dynamic range, high fidelity broadband imaging spectropolarimetry along with a sub-second time resolution and arcsecond-level angular resolution. We recommend to prioritize the implementation of such a ground-based instrument within this decade. We also call for facilitating multi-wavelength, multi-messenger observations and advanced numerical modeling in order to achieve a comprehensive understanding of the ``system science'' of solar flares and eruptions.

\newpage

\vspace{-2cm}

\section{Introduction}\label{sec:intro}

\pagenumbering{arabic}
\setcounter{page}{1}

Magnetic reconnection is a fundamental process that powers explosive energy release events on the Sun, in the heliosphere, and in many other astrophysical and space plasma systems. 
It is also a key mechanism for accelerating charged particles to high energies \citep{1997JGR...10214631M,2011SSRv..159..357Z,Li2021}, and might be responsible for heating the solar and stellar coronae to multi-million degrees \citep{1988ApJ...330..474P}. By virtue of their proximity, solar flares and solar eruptive events serve as an excellent laboratory to study the magnetic-reconnection-driven energy release and the subsequent energy conversion processes in spatiotemporal details not possible elsewhere via remote-sensing observations.

\begin{figure}[!ht]
{\includegraphics[width=1.0\textwidth]{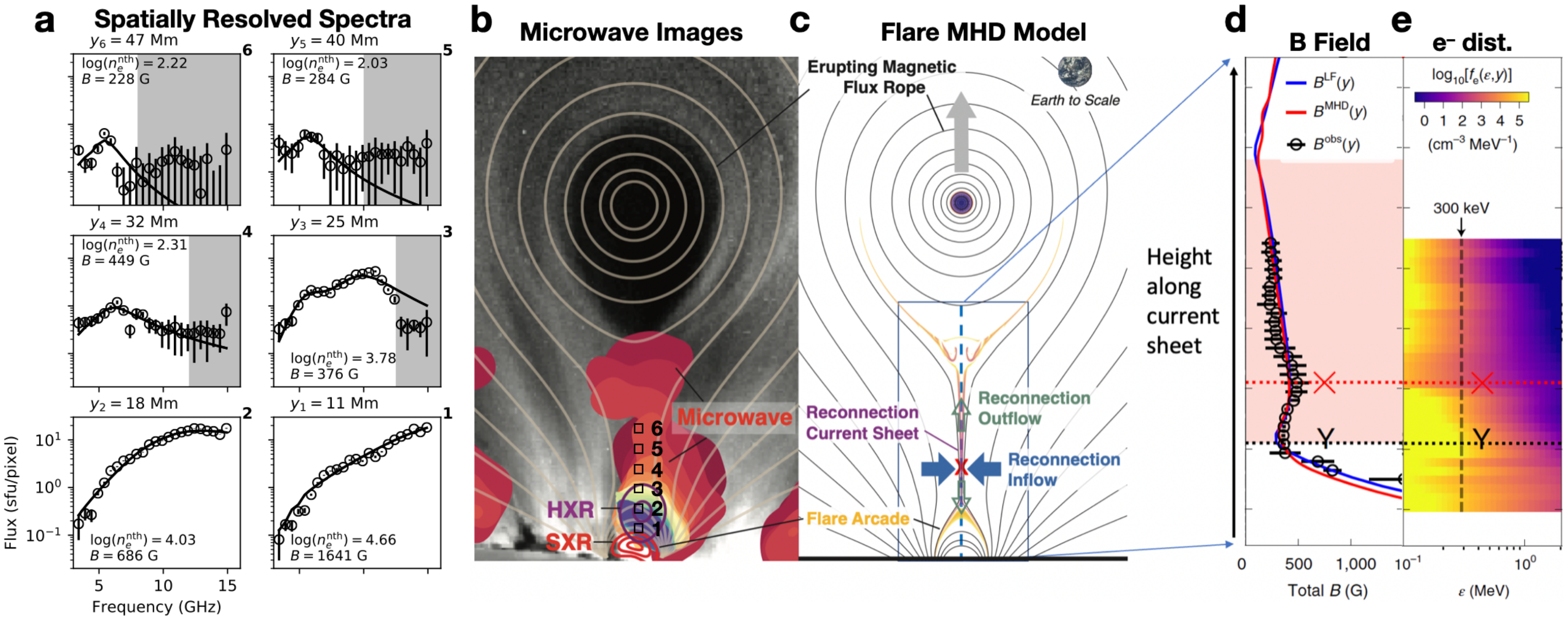}}
{\caption{\fontsize{11}{13}\selectfont Multi-wavelength observations of an eruptive solar flare event on 2017 Sept. 10. (a) Examples of spatially resolved EOVSA microwave spectra from selected locations along a large-scale reconnection current sheet (black boxes in (b)). (b) Microwave (red to blue for increasing frequency) and RHESSI hard X-ray (HXR; purple) sources from energetic electrons in the current sheet region. (c) A matching MHD model in the framework of the standard scenario. (d) Spatial variation of the magnetic field strength along the current sheet derived from the microwave data (black), which matches the model predictions (red/blue). (e) Spatial--energy distribution of energetic electrons along the current sheet. (From \citealt{2020NatAs...4.1140C}.)}\label{fig:eovsa}}
\end{figure}

For decades, we have observed solar flares with ever-increasing spatial, temporal, and spectral resolution at multiple wavelengths. Models now exist to account for a variety of phenomenological aspects of flares under different geometries. One of the most well-known is the CSHKP model 
(also referred to as the \textit{standard model} of eruptive solar flares), which involves energy release within a large-scale reconnection current sheet beneath an erupting magnetic flux rope (Fig. \ref{fig:eovsa}). 

However, until very recently, our knowledge of magnetic reconnection on the Sun was largely limited to what we could infer indirectly from observations of the morphology and dynamics of the newly reconnected magnetic loops populated with thermal plasma bright in EUV/soft X-ray emissions \citep{forbes1996,fletcher2011} or from the locations and evolution of coronal hard X-ray sources \citep{sui2003,veronig2006,krucker2010}. Now, thanks to recent advances in radio frequency (RF) and digital technology, radio observations have evolved from Fourier synthesis imaging at a limited number of frequency channels or total-power dynamic spectroscopy to the ability to do both simultaneously---a technique known as \textit{dynamic radio imaging spectroscopy}. This technique offers radio imaging with simultaneously high spectral and temporal resolution over a large number of contiguous frequency channels. Radio observations of this kind have begun to provide unique insights into the coronal magnetic field in the flaring volume and nonthermal electrons accelerated at or in close proximity to the reconnection region (Figs. \ref{fig:eovsa}(d) and (e)).

This white paper emphasizes exciting opportunities enabled by the new radio observing technique for studying \textit{magnetic reconnection and the associated magnetic energy release processes} in solar flares and solar eruptions. We will discuss the limitations of the current facilities and how we can make major breakthroughs on this topic with FASR, a next-generation ground-based radio facility equipped with superior imaging spectropolarimetry capabilities. 

\vspace{-0.1cm}

\section{Outstanding Questions and Challenges}\label{sec:questions}

A number of outstanding challenges exist in understanding the fundamental physical processes underlying the magnetic energy release and conversion processes in various space, astrophysical, and laboratory plasma contexts. A summary of these challenges is provided in a recent review paper by \citet{Ji2022a} and discussed in a white paper by \citeauthor{Ji2022} Here we outline two fundamental questions pertinent to solar flares/eruptions:

\begin{itemize}[itemsep=0.2pt,topsep=0pt]
\item  \textbf{\fontsize{13}{16}\selectfont Where and how does magnetic reconnection and energy release occur?} More specifically, what is the global magnetic and plasma context in which the reconnection occurs? 
To date, properties of magnetic reconnection has often been inferred \textit{indirectly} from the geometry or dynamics of newly reconnected magnetic loops populated with thermal plasma (e.g., X or cusp-shaped loops, plasmoid-like structures, and fast plasma outflows) 
or from coronal magnetic field models extrapolated from magnetograms measured at the photospheric/chromospheric level. 
Yet the precise magnetic structure at and around each reconnection site is largely unknown due to, in particular, the difficulty in measuring the dynamically evolving magnetic field in those locations. 

\item  \textbf{\fontsize{13}{16}\selectfont Where and how does the energy conversion occur?} Once fast magnetic reconnection is triggered, the inflowing magnetic energy in the form of Poynting flux is quickly converted into other forms of energy: nonthermal particle acceleration, heating, turbulence and waves, and bulk flows. The sites for particle acceleration and plasma heating, for example, do not necessarily coincide with the reconnection region(s), while their locations and the mechanisms responsible remain largely unknown. 
The detailed energy partition among different energy forms following magnetic reconnection is also poorly constrained by observational data.
\end{itemize}

Resolving the outstanding questions above requires the following observational capabilities: 
\begin{enumerate}
    \item Measuring the fast-evolving coronal magnetic field at and around the energy release sites. 
    \item Tracing and measuring energetic particles throughout the flaring region and beyond. 
    \item Mapping and quantifying the flare-heated plasma in a wide range of temperatures.
\end{enumerate}
Current and future observations provide outstanding means for addressing requirements \#2 and \#3, particularly when multi-wavelength analysis and modeling are utilized (see, e.g., Fig. \ref{fig:eovsa}). We refer readers to white papers by \citeauthor{Oka2022}, \citeauthor{Christe2022}, \citeauthor{Shih2022}, and \citeauthor{Kerr2022a} for perspectives on pertinent studies with optical, (E)UV, X-ray/$\gamma$-ray observations. However, measurements that meet requirement \#1 (dynamic flare magnetic field) has remained largely elusive, which constitutes one of the major challenges for advancing our understanding to the next level. In the following, we will demonstrate how broadband radio imaging spectroscopy can, on one hand, \textbf{complement observations at other wavelengths to provide key constraints for the accelerated electrons and heated plasma}, and on the other hand, offer \textbf{\textit{unique measurements} of the highly dynamic coronal magnetic field} in and around the energy release region.

\vspace{-0.5cm}

\section{Unique Opportunities at Radio Wavelengths}
\label{sec:radio_diagnostics}

It has been well known that radio observations provide unique diagnostics thanks to the access to different types of radio emission that arise in solar flares. We refer to reviews such as \citet{1998ARA&A..36..131B} and \citet{pick2008} for more details. A brief summary is included below.
\begin{itemize}[noitemsep,topsep=2pt]
\item\textbf{Coherent radio bursts} serve as a highly sensitive means for \textit{tracing nonthermal electrons} at and beyond the reconnection sites, thereby probing important characteristics of reconnection including the location, its fragmentary nature, and the presence of reconnection-driven shocks, waves, and turbulence. 

\item\textbf{Nonthermal gyrosynchrotron radiation} can be used to \textit{measure the fast-evolving coronal magnetic field} ``illuminated'' by the flare-accelerated nonthermal electrons. It also serves as an excellent tool to trace and \textit{quantify the flare-accelerated nonthermal electrons} throughout the flaring/eruption region.

\item\textbf{Thermal gyroresonance and bremsstrahlung radiation} arise from thermal electrons interacting with the ambient plasma, and can be utilized to \textit{measure the coronal magnetic field in active regions} and to derive properties of the flaring and non-flaring thermal plasma.
\end{itemize}

While the uniqueness and potential of radio diagnostics have long been recognized, the full power of such diagnostics has not been demonstrated until very recently. One of the main obstacles has been the lack of observations that allow us to obtain spatially and temporally resolved radio \textit{brightness temperature} spectra $T_b(\nu)$ for performing subsequent spectral analysis. The $T_b(\nu)$ spectra, which are equivalent to the (specific) intensity spectra in optical, UV, and X-ray spectroscopy $I_{\lambda}(\lambda)$, is the key to accessing the time- and space-varying magnetic field and energetic electron distribution. Such a need calls for radio observations covering a broad frequency range with simultaneous spatial, spectral, and temporal resolution, a technique referred to as \textit{``broadband dynamic imaging spectroscopy.''} That is, each pixel in a snapshot 2D image will have its own high-resolution broadband radio spectrum to which we can apply spectral diagnostics. 

Similar to many other spectral regimes, this technique, sometimes referred to as (temporally resolved) \textit{``integral field spectroscopy''} in the EUV and X-ray domain, is often regarded as the ``ultimate goal'' of solar observing (c.f., \citealt{2021FrASS...8...50Y}). To approach simultaneous 2D imaging and spectroscopy at shorter wavelengths, however, compromises often need to be made --- such as scanning one or more slits across the field of view or using the technique of ``overlappograms'' to disperse the image onto the detector --- at the expense of reduced time cadence or spectral purity. On the other hand, at radio wavelengths, the development of fast digital samplers ($>$1 Gsa/s), wide-band receivers ($\Delta \nu/\nu>10$:1), and fast digital correlators have allowed wide-band, Fourier-transform-based spectroscopy and 2D imaging at the same time, but without the spectral--spatial confusion expected from, e.g., multi-slit UV/EUV imaging spectrographs and overlappograms.

This revolutionary technique was not realized for solar radio studies until the beginning of the last decade  \citep{2013ApJ...763L..21C,2014A&A...568A..67M}. Today, with the successful commissioning of a number of radio facilities including the general-purpose VLA \citep{Perley2011}, LOFAR \citep{lofar2013}, the Murchison Widefield Array (MWA; \citealt{MWA2013}), and particularly, solar-dedicated arrays such as the Mingantu Spectral Radioheliograph (MUSER; \citealt{MUSER2021}), the Siberian Radioheliograph (SRH; \citealt{SRH2020}), and the Expanded Owens Valley Solar Array (EOVSA; \citealt{2018ApJ...863...83G}), science with this transformative technique has begun to flourish. Outstanding examples relevant to flare reconnection and solar eruptions include tracing fast electron beams \citep{2017ApJ...851..151M,2018ApJ...866...62C}, mapping shock fronts \citep{Carley2013,2015Sci...350.1238C,Morosan2019}, and measuring flaring magnetic fields and accelerated nonthermal electrons \citep{2020NatAs...4.1140C,2020Sci...367..278F,fleishman2022}.

Fig. \ref{fig:eovsa} shows one  example of EOVSA observations of an X8.2 GOES-class eruptive solar flare occurred on 2017 September 10 \citep{2020NatAs...4.1140C}. The imaging spectroscopy capability of EOVSA in 3.4--18 GHz (at that time) allowed us to obtain spatially resolved radio spectra (panel (a)) which, in turn, were used to derive parameters of the magnetic field and energetic electron distribution along the large-scale reconnection current sheet. These measurements enable direct comparison with model predictions to identify the magnetic nature of the reconnection sites, and moreover, used to reveal the key role of a ``magnetic bottle,'' located at the bottom of the current sheet above the looptop, in accelerating and confining the energetic electrons. More recent studies of this flare have uncovered that the electron acceleration in this region is highly efficient.  \citet{fleishman2022} reported that virtually all available source electrons there were energized to nonthermal energies. 

This example showcases one of the many diagnostics of magnetic energy release and conversion processes enabled by this new technique, particularly when complemented by other multi-wavelength data and numerical modeling. However, \textit{EOVSA} is only a small array with merely 13 antenna elements. Its angular resolution is limited to $\sim$60$''$/GHz and its snapshot image dynamic range is limited to a few $\times$10:1. Thus \textbf{most EOVSA diagnostics are limited to the regions nearest the brightest source and the intervals around the flare peak}. 

\begin{figure}[!ht]
\floatbox[{\capbeside\thisfloatsetup{capbesideposition={right,top},capbesidewidth=4cm}}]{figure}[\FBwidth]
{\includegraphics[width=0.55\textwidth]{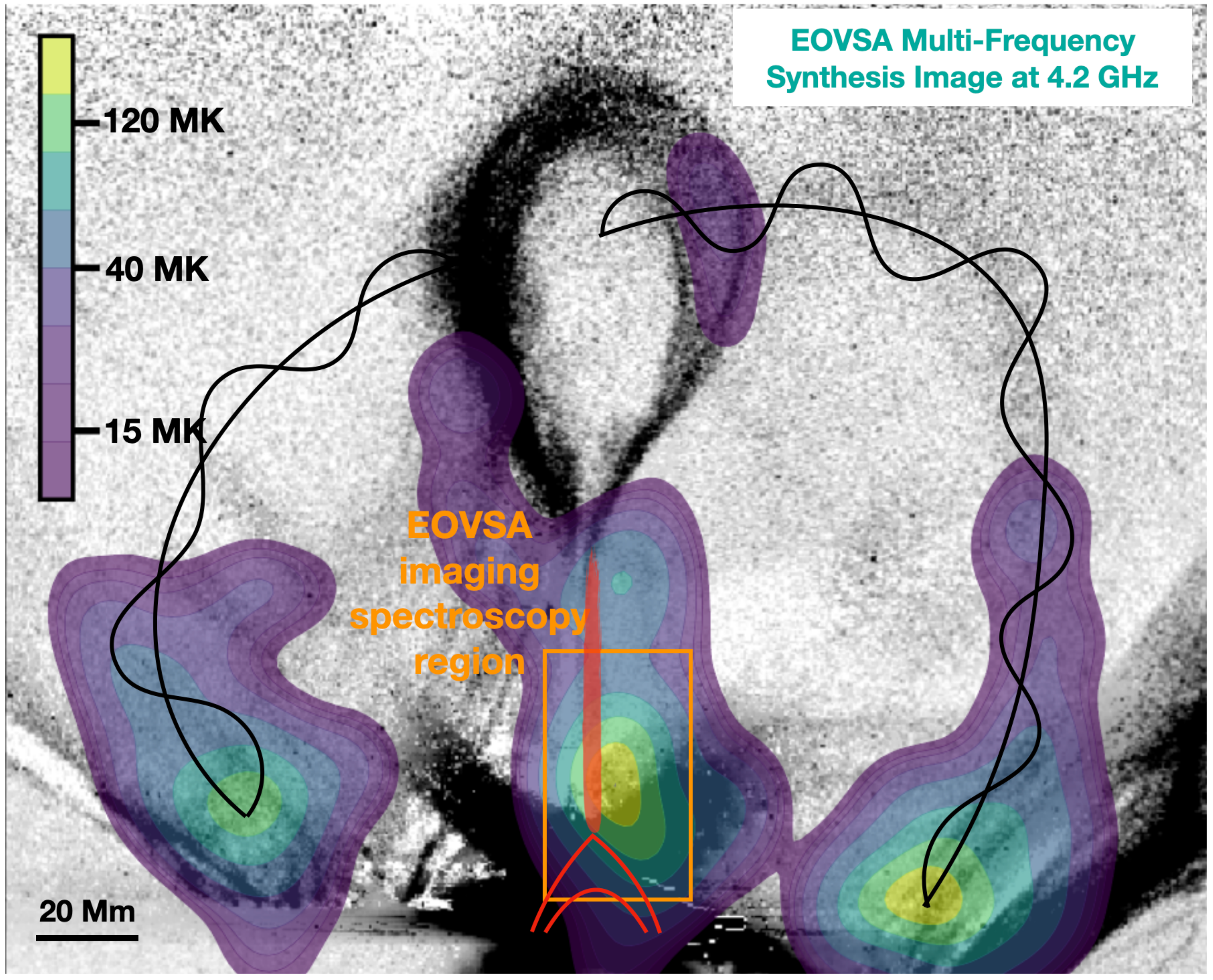}}
{\caption{\fontsize{11}{13}\selectfont EOVSA has revealed that energetic electrons have access to an extremely broad flaring region, yet spatially resolved spectral analysis is limited only to the brightest region (orange box). FASR will extend the imaging spectroscopy diagnostics to a much broader region throughout the flare and eruption volume. (From \citealt{2020ApJ...895L..50C}.)}\label{fig:spec_access}}
\end{figure}

We have long recognized that large solar eruptive events draw energy from a much broader coronal region than where the energy is ultimately released and subsequently transported in an even greater volume. Therefore, in order to obtain a complete picture of flare/eruption energy release and conversion, \textbf{\textit{it is critical to measure the time-evolving magnetic field and flare-energized electron distribution in a much broader area, preferably throughout the entire flare and eruption region}} (see Fig. \ref{fig:spec_access} for a demonstration of the large volume occupied by energetic electrons, well outside the flare arcade). In addition, a comprehensive understanding of such requires detailed analysis of events with different sizes and geometry. To this end, \textbf{\textit{it is compulsory to expand the imaging spectroscopy analysis to weaker but much more frequent events.}} This requirement also calls for an instrument with an orders-of-magnitude increase in image dynamic range, sensitivity, and a much cleaner point-spread function (PSF, or ``synthesized beam'' in radio astronomy; see white paper by \citealt{Mondal2022}). 

Last but not least, \textbf{\textit{accurate radio polarimetry is imperative to recover the angle of the magnetic field vector with respect to the line of sight}}---a much needed, but missing, capability. For example, for events with a suitable viewing geometry, it can be used to derive the \textbf{guide field component}, which is believed to play a key role in the energy storage, release, and particle acceleration and heating processes \citep{Dahlin2014,Arnold2021}.  \textbf{These coronal magnetic field measurements are not limited by photon-counting statistics and can be made at an extremely fast, sub-second cadence}. Such measurements are, on one hand, a \underline{\textit{unique}} means to study the \underline{\textit{fast-evolving}} eruption energy release, and on the other hand, perfectly complement the measurements of the flare context using novel optical/IR/EUV spectropolarimetry expected from DKIST and the proposed COSMO concept (see white paper by \citealt{Tomczyk2022}).

\vspace{-0.5cm}

\section{New Frontiers with a Next Generation Solar Radio Facility}
\begin{figure}[!ht]
{\includegraphics[width=1.0\textwidth]{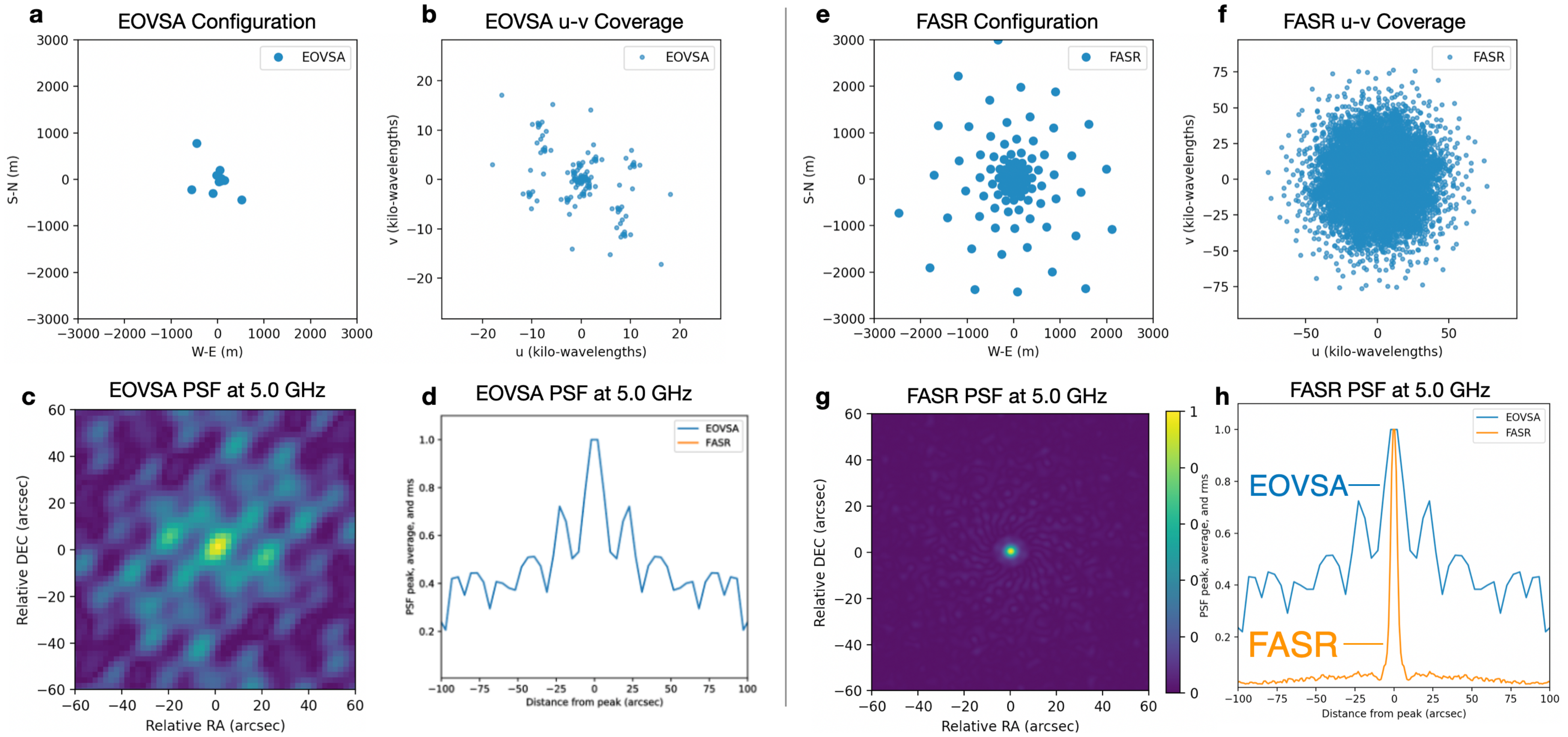}}
{\caption{\fontsize{11}{13}\selectfont FASR has an extremely clean PSF thanks to its much denser u-v coverage brought by its two orders of magnitude increase in baselines compared to EOVSA. The left and right panels show, respectively, the antenna configuration, u-v coverage, PSF image, and 1-D representation of the maximum PSF side-lobe level vs. distance for EOVSA and FASR. }\label{fig:fasr_psf}}
\end{figure}

\vspace{-0.2cm}
Major breakthroughs on this topic in the next decade and beyond call for \textbf{a next generation solar radio instrument that can perform broadband dynamic imaging spectropolarimetry with high sensitivity, high image dynamic range, high image fidelity, and high resolution}. \textbf{The \textit{Frequency Agile Solar Radiotelescope} (FASR) concept \citep{Gary2022c} constitutes such a next-generation solar radio instrument.} An overview of FASR and the broad breadth of science topics on which it can make major advances are found in another white paper by \citeauthor{Gary2022c} Briefly, as a mid-scale ($<$\$100M), ground-based community instrument, the core of the FASR concept lies in its ultra-wide bandwidth and fast digital correlators, and its orders of magnitude denser u-v coverage in a 4--5 times larger footprint compared to EOVSA. The expanded bandwidth enables broadband imaging spectroscopy across the entire $\sim$0.2--20 GHz with a sub-second time resolution (potentially down to milliseconds in a ``bursty'' mode), and the $>$2 order of magnitude increase in the number of baselines results in an improvement of the imaging dynamic range by at least two orders of magnitude (from a few$\times$10:1 for EOVSA to several $\times$1,000:1) thanks to its extremely clean PSF. FASR's superior PSF performance is shown in Fig. \ref{fig:fasr_psf}: an extremely low side-lobe level of only a few percent (compared to EOVSA's $\sim$70\%), and an angular resolution 4--5 times better than EOVSA, achieving a 1$''$ resolution at 20 GHz. The high dynamic range and angular resolution of FASR are approaching those of EUV/X-ray instruments equipped with direct focusing optics. 

\begin{figure}[!ht]
{\includegraphics[width=1.0\textwidth]{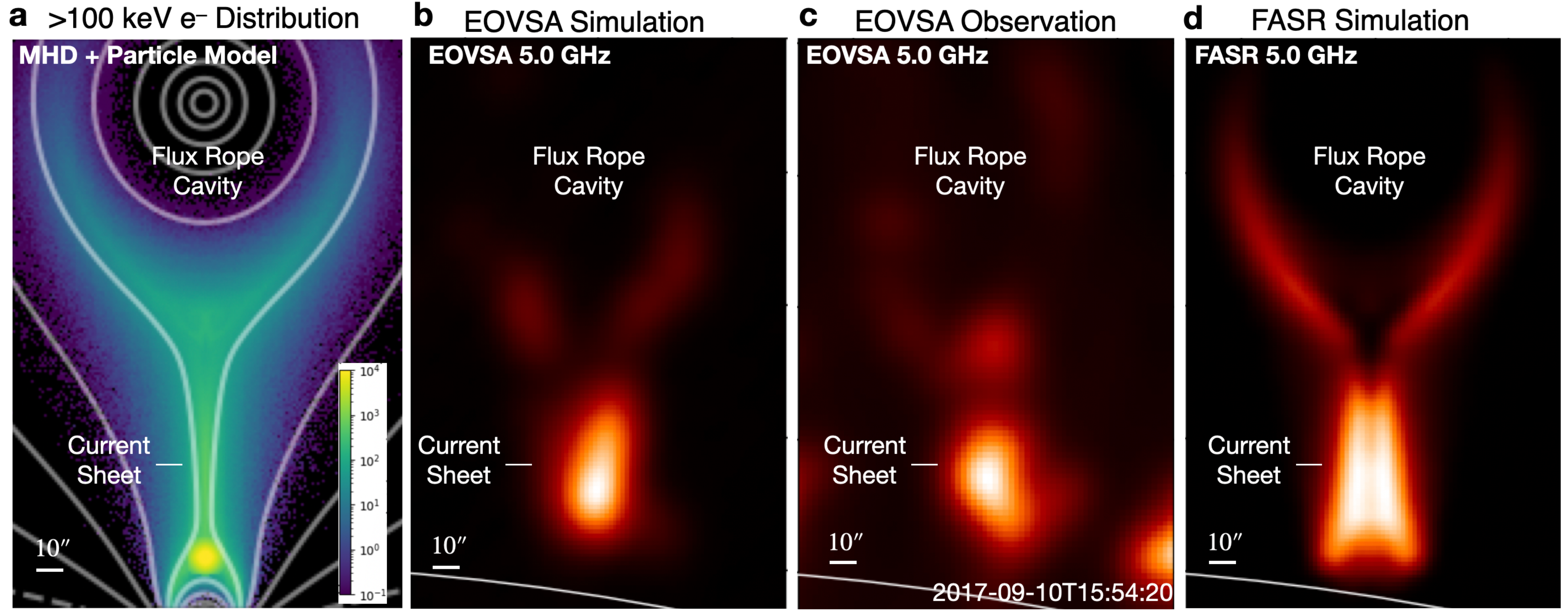}}
{\caption{\fontsize{11}{13}\selectfont FASR's superior imaging performance. The input model is made from emission maps calculated with fast gyrosynchrotron codes \citep{Kuznetsov2021} based on the output of the data-constrained MHD and particle flare model shown in (a), using the methods discussed in \citealt{LiX2022}. The model is then ``observed'' with EOVSA/FASR array configurations to produce complex visibilities using CASA's \texttt{simobserve} task, which are subsequently Fourier transformed and deconvolved with the multi-scale \texttt{CLEAN} algorithm to simulate realistic images. The simulated EOVSA image in (b) closely resembles the actual observed EOVSA image at 5 GHz shown in (c) and in Fig. \ref{fig:spec_access}. The simulated FASR image from the same model is shown in (d), demonstrating its superior imaging performance. }\label{fig:fasr_imaging}}
\end{figure}

\begin{figure}[!ht]
\floatbox[{\capbeside\thisfloatsetup{capbesideposition={right,center},capbesidewidth=4.7cm}}]{figure}[\FBwidth]
{\includegraphics[width=0.6\textwidth]{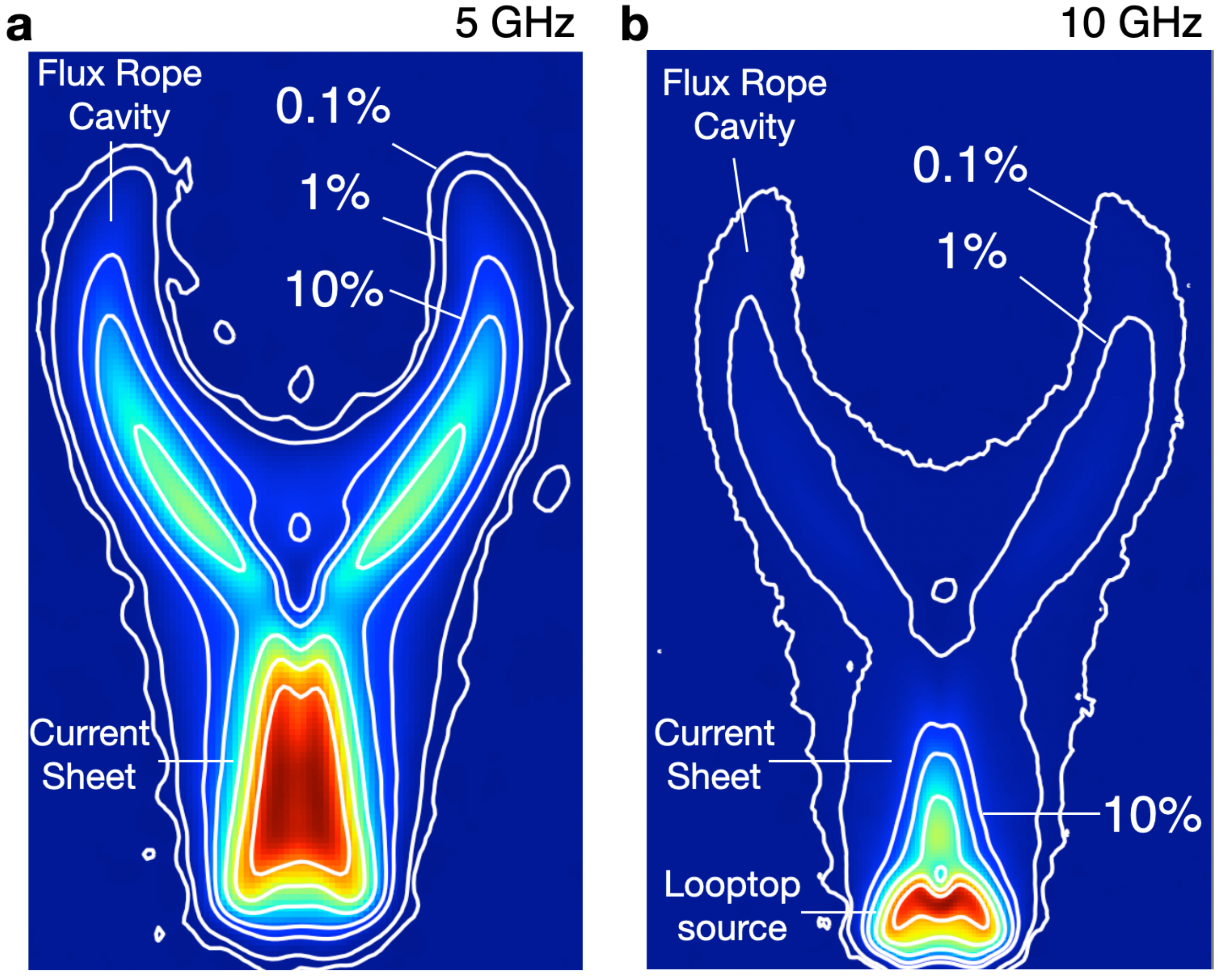}}
{\caption{\fontsize{11}{13}\selectfont FASR can achieve an ultra high dynamic range of $>$1000:1 at all frequencies, so all regions enclosed within the 0.1\% contour have sufficient SNR to perform spatially resolved spectroscopy. Example images based on the same model shown in Fig. \ref{fig:fasr_imaging}(a) are displayed for 5 GHz and 10 GHz, respectively. Note at higher frequencies the emission becomes dominated by the looptop source, yet FASR can still recover the faint emission from the current sheet region all the way into the flux rope cavity.}\label{fig:fasr_hdr}}
\vspace{-0.3cm}
\end{figure}

Fig. \ref{fig:fasr_imaging} demonstrates the superior imaging performance of FASR, based on realistic MHD and macroscopic particle simulations of the 2017 September 10 flare \citep{2020NatAs...4.1140C,LiX2022}. In agreement with the observations shown in Fig. \ref{fig:spec_access}, nonthermal electrons in the simulations have access to an extensive region at and beyond the heart of the flare/eruption and produce microwave emission (panel (a)). FASR's high fidelity, high resolution imaging will reveal unprecedented fine-scale features of all the faint radio sources, from the reconnection current sheet to the magnetic flux rope cavity.


Perhaps more profoundly, our simulations show that FASR can achieve an ultra-high dynamic range of $>$1000:1 at all frequencies (see Fig. \ref{fig:fasr_hdr}), indicating that \textbf{all regions within the 0.1\% contour are now accessible to detailed imaging spectroscopy!} Combining such a focusing-optics-like imaging performance with its full-disk ``integral field spectroscopy'' capability afforded by the radio Fourier-synthesis technique, \textbf{a high-quality spectrum can be made for any location in the extended flare/eruption region for subsequent spectral analysis}. This will yield accurate measurements of key local physical parameters including the fast-evolving coronal magnetic field, flare-accelerated nonthermal electrons, and flare-heated plasma over the entire core of the eruptive flare. Such a superior spectral imaging capability will open up completely new windows for flare diagnostics, enabling breakthrough science in the decades to come.

\vspace{-0.5cm}

\section{Concluding Remarks}

\vspace{-0.2cm}
To summarize, the outstanding capability of diagnosing key solar flare/eruption parameters with spatially resolved radio imaging spectroscopy, complemented by multi-wavelength observations and modeling, has already been demonstrated (see Sec. \ref{sec:radio_diagnostics}). The next revolution calls for prioritizing a solar-dedicated radio facility such as FASR, as well as facilitating coordinated, comprehensive imaging and spectroscopy observations and numerical modeling. In order to advance the fundamental science discussed in this white paper, our primary strategic recommendation is to:
\begin{itemize}
    \item Prioritize a FASR-like next-generation solar radio instrument that can perform broadband dynamic imaging spectropolarimetry in $\sim$0.2--20 GHz with high sensitivity, high image dynamic range, high image fidelity, and down to arcsecond resolution. 
\end{itemize}
Additionally, we recommend to:
\begin{itemize}
    \item Facilitate multi-wavelength, multi-messenger observations to achieve a comprehensive understanding of the ``system science'' of solar flares/eruptions. Achieving this goal requires multi-agency coordination to support both ground-based and space-borne facilities. 
    \vspace{-1.ex}
    \item Develop global-scale, data-constrained, 3D flare/eruption models and advanced visualization techniques to allow direct comparisons with multi-wavelength observations, so as to enable critical examination of key theories (see white paper by \citealt{Allred2022}).
\end{itemize}

\bibliography{reconnection}

\bibliographystyle{aasjournal_bc}

\end{document}